\documentclass[12pt]{article}
\usepackage{graphicx}
\usepackage{amssymb}
\usepackage{amsmath}
\newlength{\extraspace}
\newlength{\extraspaces}
\newcommand{\be}{\begin{equation}
\addtolength{\abovedisplayskip}{\extraspaces}
\addtolength{\belowdisplayskip}{\extraspaces}
\addtolength{\abovedisplayshortskip}{\extraspace}
\addtolength{\belowdisplayshortskip}{\extraspace}}
\newcommand{\ee}{\end{equation}}

\newcommand{\ba}{\begin{eqnarray}
\addtolength{\abovedisplayskip}{\extraspaces}
\addtolength{\belowdisplayskip}{\extraspaces}
\addtolength{\abovedisplayshortskip}{\extraspace}
\addtolength{\belowdisplayshortskip}{\extraspace}}
\newcommand{\ea}{\end{eqnarray}}

\newcommand{\newsection}[1]{
\vspace{12mm}
\pagebreak[3]
\addtocounter{section}{1}
\setcounter{equation}{0}
\setcounter{subsection}{0}
\noindent{\bf \thesection. #1}
\nopagebreak
\medskip
\nopagebreak}

\newcommand{\newsubsection}[1]{
\vspace{0.8cm}
\pagebreak[3]
\addtocounter{subsection}{1}
\noindent{\it \thesubsection. #1}
\nopagebreak
\vspace{2mm}
\nopagebreak}

\newcounter{saveeqn}

\newcommand{\dif}{\mathrm{d}}
\newcommand{\me}{\mathrm{e}}

\begin{document}
\addtolength{\baselineskip}{1.5mm}

\thispagestyle{empty}
\begin{flushright}

\end{flushright}
\vbox{}
\vspace{2cm}

\begin{center}
{\LARGE{Rotating black rings on Taub-NUT
        }}\\[16mm]
{Yu Chen$^1$~~and~~Edward Teo$^{1,2}$}
\\[6mm]
$^1${\it Department of Physics,
National University of Singapore, 
Singapore 119260}\\[5mm]
$^2${\it Centre for Gravitational Physics, College of Physical and Mathematical Sciences,\\[1mm]
The Australian National University, Canberra ACT 0200, Australia}\\[15mm]

\end{center}
\vspace{2cm}

\centerline{\bf Abstract}
\bigskip
\noindent
In this paper, we construct new solutions describing rotating black rings on Taub-NUT using the inverse-scattering method. These are five-dimensional vacuum space-times, generalising the Emparan--Reall and extremal Pomeransky--Sen'kov black rings to a Taub-NUT background space. When reduced to four dimensions in Kaluza--Klein theory, these solutions describe (possibly rotating) electrically charged black holes in superposition with a finitely separated magnetic monopole. Various properties of these solutions are studied, from both a five- and four-dimensional perspective.


\newpage
\newsection{Introduction}

Black holes in higher dimensions have attracted much attention in the past decade, although most of the progress has been confined to five dimensions. One catalyst for this was the discovery of a rotating black ring solution in five dimensions by Emparan and Reall \cite{Emparan:2001}. Unlike the five-dimensional Myers--Perry black hole \cite{Myers:1986}, the black ring has a non-spherical event-horizon topology. Instead, it has a ring topology $S^1\times S^2$, with rotation in the $S^1$ direction. This rotation creates a centrifugal force which opposes its self-gravity. Due to an imbalance of these two forces, there is a conical singularity in the space-time to stabilise the black ring. However, for a certain value of the angular-momentum parameter, the forces balance exactly and there is no conical singularity present. The black ring is then completely regular outside the event horizon. 

Now, black rings can also rotate in the azimuthal direction of the $S^2$. A solution describing an $S^2$-rotating black ring, without an $S^1$ rotation, was found in \cite{Mishima:2005id,Figueras:2005zp}. This solution necessarily contains a conical singularity, as there is no centrifugal force in this case to balance the self-gravity of the black ring. The real breakthrough, however, came with the derivation by Pomeransky and Sen'kov \cite{Pomeransky:2006} of a black ring rotating in {\it both\/} the $S^1$ and $S^2$ directions, and whose $S^1$ rotation has been tuned to ensure that there is no conical singularity in the space-time. The most general doubly rotating black ring, which in general has a conical singularity, was derived in \cite{Morisawa:2007di,Chen:2011jb}.

All the above-mentioned solutions have backgrounds that are simply flat space. By this we mean that when the black hole/ring is removed from the space-time---for example by setting its mass to zero---the resulting background is nothing but a direct product of four-dimensional flat space and a flat time direction. More recently, five-dimensional black holes/rings on backgrounds other than flat space have been considered (see, e.g., \cite{Chen:2010ih} for a survey). Of these, Taub-NUT space \cite{Newman:1963,Hawking:1976} is perhaps the simplest non-trivial possibility. These space-times are asymptotically locally flat, i.e., they asymptotically approach a twisted $S^1$ bundle over four-dimensional Minkowski space-time. Such space-times are of special interest in Kaluza--Klein theory, where the fifth dimension is assumed to be compactified into a circle. At distances much smaller than the radius of the $S^1$---near the so-called nut---the space-times become isometric to five-dimensional Minkowski space-time.

Black rings on a Taub-NUT background were first constructed within the context of a supersymmetric theory \cite{Elvang:2005,Gaiotto:2005,Bena:2005}. In these solutions, we can view the black-ring horizon as wrapping the $S^1$ fibre of the Taub-NUT space. When the radius of the ring is made a tunable parameter, the black ring effectively interpolates between a five-dimensional solution (for small radii) and a four-dimensional one (for large radii). This fact has been used, for example, to find a connection between four- and five-dimensional black-hole partition functions \cite{Gaiotto:2005gf,Gaiotto:2005}.

In this paper, we are primarily interested in black rings on Taub-NUT in vacuum Einstein gravity. It turns out to be somewhat more difficult to construct such solutions than in a supersymmetric theory, as there is no longer a linear structure to the field equations. For a flat-space background, the inverse-scattering method (ISM) \cite{Belinski:2001,Pomeransky:2005sj} has proven to be a fruitful way to generate new black-ring solutions starting from a suitable seed solution (see, e.g., \cite{Iguchi:2011qi,Chen:2011jb}). However, the challenge has been to generate new solutions describing black rings on Taub-NUT. One way to do so was developed in \cite{Giusto:2007fx,Ford:2007}, which uses an $SL(3,\mathbb{R})$ transformation to turn a black ring on flat space to one on Taub-NUT. Thus, black rings on Taub-NUT can be generated using a two-step process beginning with the ISM, and followed by an $SL(3,\mathbb{R})$ transformation.

Using this technique, the static black ring on Taub-NUT was first constructed by Ford et al.~\cite{Ford:2007}. Like the static black ring on flat space \cite{Emparan:2001wk}, this solution is not completely regular because of the presence of a conical singularity in the space-time, although it can be eliminated by introducing rotation.
A rotating black ring on Taub-NUT was subsequently constructed by Camps et al.~\cite{Camps:2008}. It has the property that it rotates in both the $S^1$ and $S^2$ directions of the ring, although the two rotations are not independent: they are governed by a single rotational parameter. The space-time is free of conical singularities for a specific choice of this parameter. 

Rotating black rings on Taub-NUT have an interesting interpretation in Kaluza--Klein theory. The black ring reduces to a (possibly rotating) electrically charged black hole, while the nut reduces to the well-known Kaluza--Klein monopole \cite{Gross:1983,Sorkin:1983}. Thus the reduced four-dimensional solution describes a black hole a finite distance apart from the monopole in an asymptotically flat space-time, with a conical singularity in general connecting them. Even if the black hole were not rotating, it is remarkable that the space-time contains angular momentum arising from the crossed electric and magnetic fields of the black hole and monopole, respectively. Such a phenomenon also occurs in classical electrodynamics (see, e.g., \cite{Jackson}).

It is also possible to embed this Kaluza--Klein system in Type IIA string theory \cite{Camps:2008}. The magnetic monopole uplifts to a D6-brane in 10 dimensions, while the black hole uplifts to a set of D0-branes. Thus, the solution of \cite{Camps:2008} would describe a (non-supersymmetric) system of D0-branes in the presence of a D6-brane. In general, the D0-branes are thermally excited, and the zero-temperature limit can be obtained by taking the extremal limit of the black hole. However, a conical singularity is necessarily present in this limit; it cannot be eliminated for any choice of the rotational parameter.

Now, the solution of Camps et al.~\cite{Camps:2008} is not the most general rotating black ring on Taub-NUT possible. Such a solution would have independent rotational parameters along the $S^1$ and $S^2$ directions. Like in \cite{Camps:2008}, the $S^1$ rotation can be tuned to eliminate the conical singularity. However, such a solution would contain a second rotational parameter, which could, for example, be used to take the extremal limit of the black ring. This would enable one to obtain an extremal black ring on Taub-NUT that is regular outside the event horizon, unlike the one considered in \cite{Camps:2008}. It would be of interest to try to construct the most general doubly rotating black ring on Taub-NUT, although based on our experience for the doubly rotating black ring on flat space \cite{Chen:2011jb}, this solution is likely to be very complicated.

The aim of this paper is to present new vacuum solutions describing rotating black rings on Taub-NUT, representing special cases of this most general solution. We construct these solutions using only the ISM. This has an advantage over the $SL(3,\mathbb{R})$ generating technique, in that one has a better handle on which rotation to turn on. Indeed, we are able to generate an $S^1$-rotating black ring on Taub-NUT that, unlike the solution found in \cite{Camps:2008}, has {\it no\/} $S^2$ rotation, and therefore can be regarded as the natural generalisation of the Emparan--Reall black ring. Moreover, the ISM seems able to give more manageable expressions for the metric coefficients than the $SL(3,\mathbb{R})$ generating technique; it turns out that our solution takes a much simpler form than the one in \cite{Camps:2008}. The conical singularity in the space-time can be eliminated for a specific choice of the rotational parameter. When reduced to four dimensions, this solution describes a static, electrically charged black hole in equilibrium with a magnetic monopole. However, as mentioned above, the space-time will still possess an angular momentum coming from the crossed electric and magnetic fields.

We then generalise our solution to include $S^2$ rotation. This black ring on Taub-NUT has two rotational parameters, although we will fix one of them to eliminate the conical singularity in the space-time. Our solution is therefore completely regular outside the event horizon, and can be regarded as the natural generalisation of the Pomeransky--Sen'kov black ring \cite{Pomeransky:2006}. Unfortunately, its form is still sufficiently complicated that we will only present the extremal limit of this solution. It turns out that there are two different classes arising when the extremal limit is taken, differing in the sign of the $S^2$ rotation. When reduced to four dimensions, our solutions describe a rotating, electrically charged extremal black hole in equilibrium with a magnetic monopole. The total angular momentum of the space-time now includes contributions from both the crossed electric and magnetic fields, and the rotation of the black hole (which can be thought of as opposite for the two classes). Such solutions may have potential applications in string theory.

The organisation of this paper is as follows: In Sec.~2, the ISM construction of our solutions is described. In particular, we explain how the ISM can be used to introduce NUT charge into the space-time. In Sec.~3, the Emparan--Reall black ring on Taub-NUT is presented, and its properties analysed from both a five- and four-dimensional perspective. In Sec.~4, two classes of solutions, both describing an extremal Pomeransky--Sen'kov black ring on Taub-NUT, are presented. Their properties are analysed, with particular attention paid to the differences between the two classes. Readers interested only in the solutions and their properties may skip directly to Sec.~3 or 4.

\newsection{ISM construction}
\label{sec_ISM}

The inverse-scattering method has proven to be a powerful yet practical way to generate new solutions of Einstein gravity, given a known simpler seed solution. It can be used, for example, to generate the Kerr-NUT solution starting from the Schwarzschild solution \cite{Belinski:2001}. In recent years, it has been realised that the ISM can also be applied to five-dimensional Einstein gravity; for example, it can be used to generate the Myers--Perry solution starting from the Schwarzschild--Tangherlini solution \cite{Pomeransky:2005sj}. In fact, it has been used with much success to generate new black-hole solutions in five dimensions, such as the doubly rotating black ring \cite{Pomeransky:2006,Chen:2011jb}, multiple-horizon solutions like the black Saturn \cite{Elvang:2007rd} and bi-ring \cite{Elvang:2007hs}, as well as black holes on non-trivial gravitational instanton backgrounds \cite{Chen:2010ih}. A recent review of some of this progress can be found in \cite{Iguchi:2011qi}.

Our construction of the rotating black ring on Taub-NUT solution is based on the following two facts:

(i) The self-dual Taub-NUT instanton can be obtained as a special limit of the Euclidean Kerr-NUT metric. The latter metric has three parameters $m$, $a$ and $n$ (we are following the conventions of, say \cite{Chen:2010zu}, here), and setting $n=\pm m$ gives the self-dual Taub-NUT instanton. In terms of the rod structure of the Euclidean Kerr-NUT metric \cite{Chen:2010zu}, this limit corresponds to joining up the rod representing the Euclidean horizon with either the left or right semi-infinite rod.
This means that the directions of the two rods become identical in this limit, so that the turning point at which they meet becomes a ``phantom point'' and is effectively removed from the rod structure.
We note that $a$ actually becomes a redundant parameter in the limiting self-dual Taub-NUT instanton, so we have the freedom to fix it as we want for convenience. Note that, however, if $a=0$, the $n=\pm m$ limit of the resulting non-self-dual Taub-NUT solution corresponds to shrinking the length of (Euclidean) horizon-rod to a point in the rod structure, making this limit subtler to carry out in practice.

(ii) The Euclidean Kerr-NUT solution can be straightforwardly constructed using the ISM: one just needs to perform a two-soliton transformation on the Euclidean Schwarzschild seed-solution.

When combined, the above two observations allow us to generate the self-dual Taub-NUT instanton using ISM straightforwardly. Moreover, we can use the freedom in choosing the value of the parameter $a$ in (i) above to make the effect of one of the two solitons involved in (ii) trivial. In fact, we can directly perform a one-soliton transformation on the Euclidean Schwarzschild solution to generate the self-dual Taub-NUT instanton. To add a black ring to the Taub-NUT background space is then an easy matter, since the procedure is already known in the literature. In the rest of this section, we will present the full construction of rotating black rings on Taub-NUT. Some familiarity with the ISM, at the level of, say \cite{Pomeransky:2005sj}, will be assumed of the reader.

\begin{figure}[t]
\begin{center}
\includegraphics{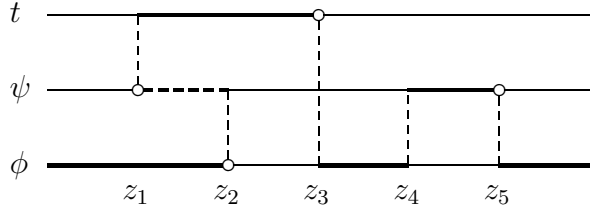}
\caption{The rod sources of the seed solution for the doubly rotating black ring on Taub-NUT. The thin lines denote the $z$-axis and the thick lines denote rod sources of mass $\frac{1}{2}$ per unit length along this axis. The dashed horizontal line denotes a rod source with negative mass density $-\frac{1}{2}$. Small circles represent the operations of removing solitons from the seed, each with a BZ vector having a single non-vanishing component along the coordinate that labels the $z$-axis where the circle is placed.}
\label{Seed_BR_TN}
\end{center}
\end{figure}

We start with a seed solution having the rod structure as shown in Fig.~\ref{Seed_BR_TN}. The explicit solution corresponding to this rod structure can be directly read off \cite{Emparan:2001wk} in Weyl--Papapetrou coordinates $(\rho,z)$ as
\ba
G_0&=&{\textrm{diag}}\, \bigg\{ -{\frac {{\mu_1}}{{\mu_3}}},\frac{\mu_2\mu_4}{\mu_1\mu_5}, \frac{\rho^2\mu_3\mu_5}{\mu_2\mu_4}\bigg\}\,,\cr
\me^{2\gamma_0}&=&k^2\,\frac{\mu_3 \mu_5 R_{12} R_{13} R_{14} R_{23} R_{25}^2 R_{34} R_{45}^2}{\mu_2\mu_4 R_{11} R_{22} R_{33} R_{44} R_{55} R_{15} R_{35} R_{24}^2}\,,
\ea
where $\mu_i\equiv\sqrt{\rho^2+(z-z_i)^2}-(z-z_i)$, $R_{ij}\equiv\rho^2+\mu_i\mu_j$, and $k$ is an arbitrary integration constant. Using the ISM, we then perform the following soliton transformations on the above seed:
\begin{enumerate}
  \item Remove a soliton at each of $z_1$, $z_2$, $z_3$ and $z_5$, with trivial Belinski--Zakharov (BZ) vectors $(0,1,0)$, $(0,0,1)$, $(1,0,0)$ and $(0,1,0)$, respectively;
  \item Add back a soliton at each of $z_1$, $z_2$, $z_3$ and $z_5$, with non-trivial  BZ vectors $(C_1,1,0)$, $(0,C_2,1)$, $(1,0,C_3)$ and $(0,1,C_5)$ respectively. Here, $C_1$, $C_2$, $C_3$ and $C_5$ are the new, so-called BZ parameters.
\end{enumerate}

In the first step, the act of removing a soliton at $z=z_k$ with a trivial BZ vector having a non-vanishing $a$-th component, refers to multiplying the diagonal element $(G_{0})_{aa}$ of the seed solution by a factor $-\frac{\mu_k^2}{\rho^2}$. So in the current case, after the first step, we obtain the new $G$-matrix:
\be
\tilde{G}_0= {\textrm{diag}}\,  \bigg\{\frac{\mu_1 \mu_3}{\rho^2} ,{\frac {{\mu_1}{\mu_2}{\mu_4}\mu_5}{{\rho}
^{4}}},-{\frac {{\mu_2}{\mu_3}\mu_5}{{\mu_4}}} \bigg\}\,.
\ee
The generating matrix $\tilde{\Psi}_0$ can be obtained directly by performing the following replacements to the $\tilde{G}_0$ matrix: $\mu_i\rightarrow \mu_i-\lambda$ and $\rho^2\rightarrow \rho^2-2z\lambda-\lambda^2$, where $\lambda$ is a spectral parameter. One can then easily follow \cite{Pomeransky:2005sj} to carry out the second step. In computing the vectors $m^{(k)}$, we used the same trick as was described in \cite{Chen:2011jb}. The conformal factor can also be easily calculated:
\be
\label{final_f}
\me^{2\gamma}=\me^{2\gamma_0}\frac{\det\Gamma(C_1,C_2,C_3,C_5)}{\det\Gamma(C_1=0,C_2=0,C_3=0,C_5=0)}\,,
\ee
where the matrix $\Gamma$, as defined in \cite{Pomeransky:2005sj}, is that corresponding to the second step above.

In this paper, we shall impose the normal ordering of the positions of turning points, i.e., $z_1< z_2< z_3 <z_4<z_5$. Now we can calculate the rod structure of the generated solution. We then join up Rods 1 and 2, as well as Rods 5 and 6 (all counted from the left), by requiring that Rod 1 has the same direction as Rod 2, and Rod 5 has the same direction as Rod 6. This fixes the BZ parameters $C_1$ and $C_5$ as follows:
\be
\label{C_14}
C_1=\sqrt{\frac{z_1^2z_{31}}{z_{21}z_{41}z_{51}}}\,,\qquad C_5=\sqrt{\frac{4z_{5}^4z_{53}}{z_{51}z_{54}^2}}\,,
\ee
where $z_{ij}\equiv z_i-z_j$. Note that, without loss of generality, we have chosen only the positive solutions of $C_{1}$ and $C_5$ in (\ref{C_14}). By setting these values of the BZ parameters, we effectively eliminate the turning points $z_1$ and $z_5$; this will leave the resulting solution with three genuine turning points. Note that the solutions (\ref{C_14}) do not depend on the other BZ parameters, so in hindsight, it is much easier to get them by assuming $C_{2,3}=0$.

Before proceeding further, let us make some remarks on the above ISM construction here. The negative density rod placed between $z_1$ and $z_2$, together with the soliton transformations performed at the turning points $z_1$, $z_2$ and $z_3$, is the usual procedure used to generate doubly rotating black ring solutions. It has previously been used to construct doubly rotating black rings \cite{Pomeransky:2006,Chen:2011jb} in five-dimensional asymptotically flat space-time (in which case the right semi-infinite rod is placed along the coordinate $\psi$). If we shrink the negative density rod to zero so that $z_1=z_2$ and set $C_{1,2,3}=0$, we obtain a static solution.

On the other hand, the purpose of the soliton transformation performed at $z_5$ is to add NUT charge: if we further impose $z_2=z_3$ in the static limit, and impose the value of $C_5$ as in (\ref{C_14}) so that the last two rods in Fig.~\ref{Seed_BR_TN} are joined up, we get the (one-soliton) ISM construction of the Taub-NUT background space, as described at the beginning of this section.

We are now in a position to cast the solution in C-metric-like coordinates $(x,y)$, which are related to the Weyl--Papapetrou coordinates by \cite{Harmark:2004rm}
\ba
\label{C-metric}
\rho = \frac{2\varkappa^2 \sqrt{(1-x^2)(y^2-1)(1+cx)(1+cy)}}{(x-y)^2}\,,\qquad
 z   =\frac{\varkappa^2 (1-xy) (2+cx+cy)}{(x-y)^2}\,.
\ea
We also make the following parameter redefinitions:
\ba
z_1=-\frac{2b-c(1+b)}{1-b}\,\varkappa^2,\quad z_2=-c\varkappa^2,\quad z_3=c\varkappa^2,\quad z_4=\varkappa^2,\quad z_5=a\varkappa^2.
\ea
Note that the choice of the parameter $b$ here is the same as for the Emparan--Reall black ring, in the notation of, say \cite{Harmark:2004rm}. Now since the solution after imposing (\ref{C_14}) has just three genuine turning points, its metric can be written in terms of purely algebraic expressions of $x$ and $y$. Some technical details can be found in \cite{Chen:2011jb} on how to convert the solution from Weyl--Papapetrou coordinates to C-metric-like coordinates. Although the conversion in \cite{Chen:2011jb} only involves eliminating one phantom turning point, the spirit is the same. This simplifies the solution dramatically.

Now this leaves six parameters in our solution: $\varkappa$, $a$, $b$, $c$, $C_2$ and $C_3$. Note that we have not eliminated all the singularities yet. In the following sections, we will discuss the Emparan--Reall black ring on Taub-NUT and the Pomeransky--Sen'kov black ring on Taub-NUT separately, since in the former we allow the presence of a conical singularity, while in the latter we do not.

\newsection{Emparan--Reall black ring on Taub-NUT}

The Emparan--Reall black ring on Taub-NUT can be obtained from the above six-parameter solution by setting
\be
C_2=C_3=0\,.
\ee
This means that effectively only a two-soliton transformation is needed in this construction. A linear transformation on the $G$-matrix \cite{Chen:2010zu} was then done and a particular choice of the integration constant $k$ made to bring the solution to precisely the form used in this section.

\newsubsection{The metric and rod structure}

The metric we found for the Emparan--Reall black ring on Taub-NUT can be written in the following form:
\ba
\label{ER_TN_metric}
\dif s^2&=&-\frac{K(x,y)}{H(x,y)}\,(\dif t+\omega_1\,\dif \psi+\omega_2\,\dif \phi)^2+\frac{F(x,y)}{K(x,y)}\,(\dif \psi+\omega_3\, \dif \phi)^2-\frac{4\varkappa^4G(x)G(y)H(x,y)}{(x-y)^4F(x,y)}\,\dif \phi^2\cr
&&+\frac{4\varkappa^4(1-c)^2 H(x,y)}{(a-c)(1-b)\Psi(x-y)^3}\left(\frac{\dif x^2}{G(x)}-\frac{\dif y^2}{G(y)}\right),
\ea
where
\ba
\omega_1&=&\frac{\sqrt{b(1+b)(b-c)\Psi}}{K(x,y)}(x-y)J_2(x,y)\,,\cr
\omega_2&=&\frac{\varkappa^2(a-1)(a+c)\sqrt{b(1-b^2)(b-c)}}{\sqrt{a-c}\,(x-y)K(x,y)}[(1+y)(x-y)J_1(x,y)-(1+x)(1+cx)J_3(x,y)]\,,\cr
\omega_3&=&\frac{\varkappa^2(a-1)(a+c)\sqrt{1-b}}{\sqrt{(a-c)\Psi}(x-y)F(x,y)}\Big\{2(a-1)(b-c)(a-c)(x-1)(1+by)(x+y+2)\cr
&&+(1+by)[(1-b)(a-c)(1-cx)-2(1-c)(a+bcx)]J_3(x,y)\cr
&&+(1+b)(b-c)(1+y)(x-y)J_2(x,y)\Big\}\,,
\ea
and the functions $G$, $H$, $F$, $K$, $J_1$, $J_2$, $J_3$, and the constant $\Psi$ are given by
\ba
G(x)&=&(1-x^2)(1+cx)\,,\cr
H(x,y)&=&(1-b)[(a+c)(1+bx)J_1(x,y)+(b-c)(1-x)J_2(x,y)]\,,\cr
F(x,y)&=&(1-b)[(b-c)(1-y)J_2(x,y)-(a+c)(1+cx)(1+by)(c+cxy+ax+ay)]\,,\cr
K(x,y)&=&(1-b)(a+c)(1+by)J_1(x,y)+(b-c)(1-x-b-bx+2by)J_2(x,y)\,,\cr
J_1(x,y)&=&(a-1)(1+x)(a+cx)+J_2(x,y)\,,\cr
J_2(x,y)&=&(a-c)(1-x)(1+cx)-(1+y)(a+cx)^2,\cr
J_3(x,y)&=&(a-c)(x^2+y^2-2)-c(1+x)(1+y)(x-y)\,,\cr
\Psi&=&(1-c)(1+b)+(a-1)(1-b)\,.
\ea
The four parameters in the above metric, $a$, $b$, $c$ and $\varkappa$, satisfy the constraints
\be
\label{range_ER_TN}
0\leq c\leq b<1< a\,, \qquad \varkappa>0\,.
\ee
As will be clear below, these parameters have the following interpretations: roughly speaking, $\varkappa$ sets the scale of the solution, $c$ characterises the size of the black hole, $b$ is a rotational parameter, while $a$ controls the NUT charge. The coordinates $t$, $x$ and $y$ take the range
\be
-\infty<t<\infty\,, \qquad -\frac{1}{c}\leq y\leq -1\leq x\leq 1\,.
\ee
The black hole horizon is located at $y=-\frac{1}{c}$, while physical infinity is located at $(x,y)=(-1,-1)$.

\begin{figure}[t]
\begin{center}
\includegraphics{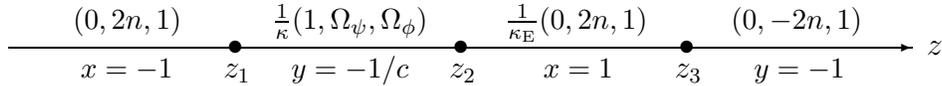}
\caption{The rod structure of the Emparan--Reall black ring on Taub-NUT, where $n$ is the NUT-charge parameter. The third turning point $z_3$ represents the location of the ``nut'', which is a fixed point of the $U(1)\times U(1)$ isometry. The black ring is completely regular outside the event horizon in the case $\kappa_{\rm E}=1$.}
\label{Figure_ER_TN}
\end{center}
\end{figure}

The solution has an obvious isometry group $\mathbb{R}\times U(1)\times U(1)$, generated by the three Killing vector fields $\frac{\partial}{\partial t}$, $\frac{\partial}{\partial \psi}$ and $\frac{\partial}{\partial \phi}$. So it allows us to apply the rod-structure formalism, following the prescription of \cite{Chen:2010zu,Chen:2010ih} (see also \cite{Harmark:2004rm,Hollands:2007aj}). In the Weyl--Papapetrou coordinates $(\rho,z)$, which are related to the above C-metric-like coordinates by (\ref{C-metric}), the rod structure has three turning points: at $(\rho=0,z=z_1\equiv-c\varkappa^2)$ or $(x=-1,y=-\frac{1}{c})$, $(\rho=0,z=z_2\equiv c\varkappa^2)$ or $(x=1,y=-\frac{1}{c})$,  and $(\rho=0,z=z_3\equiv \varkappa^2)$ or $(x=1,y=-1)$. They divide the $z$-axis into four rods:

\begin{itemize}
\item Rod 1: a semi-infinite space-like rod located at $(\rho=0, z\leq z_1)$ or $(x=-1,-\frac{1}{c}\leq  y< -1)$, with direction $\ell_1=(0,2n,1)$, where the NUT charge $n$ is given by
\be
n={\frac { {\varkappa}^{2}( a+c )  ( a-1 ) \sqrt {1-b}}{2\sqrt {(a-c)\Psi}}}\,.
\ee

\item Rod 2: a finite time-like rod located at $(\rho=0, z_1\leq z\leq z_2)$ or $(-1\leq x\leq 1,y=-\frac{1}{c})$, with direction $\ell_2=\frac{1}{\kappa}(1,\Omega_{\psi},\Omega_{\phi})$, where the surface gravity $\kappa$ and angular velocities $\Omega_\psi$ and $\Omega_\phi$ are given by
\ba
\label{ER_surface_gravity}
\kappa&=&{ \frac {\left( 1+c \right)\sqrt {(1-b)\Psi} }{4{\varkappa}^{2}
 \left( a+c \right)  \left( 1-c \right) \sqrt {cb \left( 1+b \right) }
}}
\,,\cr
\Omega_{\psi}&=&-{\frac {\sqrt {b-c} [(1+b)(1+ac)-2(a+b)-(a-2c)\Psi] }{ 2\left( 1-c \right)\left( a-c \right) \sqrt {b(1+b)\Psi}  }}\,,\cr
\Omega_{\phi}&=&-{\frac {\left( a-1 \right)\sqrt {(1-b)(b-c)}  }{ 2{\varkappa}^{2}\left( 1-c
 \right) \left( a+c
 \right)\sqrt {b(1+b)(a-c)}  }}
\,.
\ea

\item Rod 3: a finite space-like rod located at $(\rho=0, z_2\leq z\leq z_3)$ or $(x=1,-\frac{1}{c}\leq y\leq -1)$, with direction $\ell_3=\frac{1}{\kappa_{\rm E}}(0,2n,1)$, where the Euclidean surface gravity $\kappa_{\rm E}$ is defined as
\be
\label{kappa_E}
{\kappa_{\rm E}}={\frac { \left( 1+c \right) \sqrt {(a-c)\Psi}
}{ \left( 1-c \right)\left( a+c \right) \sqrt {1+b }  }}\,.
\ee

\item Rod 4: a semi-infinite space-like rod located at $(\rho=0, z\geq z_3)$ or $(-1<x\leq 1,y=-1)$, with direction $\ell_4=(0,-2n,1)$.
\end{itemize}
This rod structure is illustrated in Fig.~\ref{Figure_ER_TN}.

Here we identify \{$\ell_1$, $\ell_4$\} as the two independent $2\pi$-periodic generators of the $U(1)\times U(1)$ isometry subgroup of this solution, thus making the following identifications on the coordinates \cite{Chen:2010zu}:
\be
\label{identifications_coor}
(\psi,\phi)\rightarrow (\psi+4n\pi,\phi+2\pi)\,, \qquad (\psi,\phi)\rightarrow (\psi+8n\pi,\phi)\,.
\ee
In the rod structure, Rod 2 is time-like, so it represents an event horizon with a topology $S^1\times S^2$, which can be seen from the fact that $\ell_1\, /\hskip-2pt/\, \ell_3$. More specifically, $\ell_1$ generates the rotational symmetry of the $S^2$, while $\ell_4$ (or alternatively $4n\frac{\partial}{\partial \psi}$) generates the $S^1$. Under the above identifications, the solution (\ref{ER_TN_metric}) in general possesses a conical singularity along Rod 3. If we define the angle variable $\eta$  by $\frac{\partial}{\partial \eta}=2n\frac{\partial}{\partial \psi}+\frac{\partial}{\partial \phi}$, the conical excess along the axis represented by this rod  can then be calculated as
\be
\label{Delta_phi}
\Delta\eta=2\pi(\kappa_\mathrm{E}-1)\,.
\ee
A completely regular class, describing a regular black ring on Taub-NUT, can be obtained by imposing $\ell_1=\ell_3$, or equivalently $\kappa_{\rm E}=1$. Solving this gives
\be
\label{regularity_condition}
b={\frac {2c \left( a-{c}^{2} \right) }{   \left( 1
+{c}^{2} \right)\left( a+c \right) -2c \left( 1+c \right) }}\,.
\ee
Observe that this value of $b$ can only be achieved if $a > \frac{c(1+3c)}{1-c}$. This is compatible with $a > 1$ if $0 < c < \frac{1}{3}$. For $\frac{1}{3} < c < 1$, values of $a$ between 1 and $\frac{c(1+3c)}{1-c}$ will not give rise to any regular solutions. This result means that balance can only be achieved if the NUT charge is sufficiently large, or if the black ring is sufficiently light. We note that (\ref{regularity_condition}) reduces to the usual regularity condition $b=\frac{2c}{1+c^2}$ for the Emparan--Reall black ring in the limit $a\rightarrow \infty$.

\bigskip

\newsubsection{Various limits}

\begin{itemize}
  \item Background limit
\end{itemize}

We now show that the background space of the above solution is indeed the self-dual Taub-NUT instanton \cite{Newman:1963,Hawking:1976}. To see this, we take the limit $b,c\rightarrow 0$. After defining the new coordinates and parameter as
\be
\label{bg_transf}
r=-\frac{\varkappa^2(x+y)}{x-y}\,, \qquad \cos\theta=\frac{x^2+y^2-2}{x^2-y^2}\,,\qquad n_0=\frac{\varkappa^2(a-1)}{2}\,,
\ee
the spatial background of the solution (\ref{ER_TN_metric}) reduces to the self-dual Taub-NUT instanton in the usual form:
\be
\label{SD_TN}
\dif s^{2}=H^{-1} { \left( {\dif\psi}+2n_0\cos  \theta\,
  {\dif\phi} \right) ^{2}}+H ( {\dif r}^{2}+{r}^{2}{
{\dif\theta}}^{2}+{r}^{2} \sin ^2 \theta \,{{\dif\phi}}^{2} )\,,\quad H=1+\frac{2 n_0}{r}\,.
\ee

\begin{itemize}
  \item Static limit
\end{itemize}

This limit is taken as $b=c$, in which case both $\Omega_{\psi}$ and $\Omega_{\phi}$ vanish and we recover a static black ring on Taub-NUT. We point out that this solution is obtained using effectively a one-soliton transformation in the ISM construction. It was originally constructed by Ford et al.\ \cite{Ford:2007} in a different and much more complicated form using an $SL(3,\mathbb{R})$ generating technique. Note that in this case, the regularity condition (\ref{regularity_condition}) cannot be satisfied: a conical (strut) singularity is necessarily present to prevent the black ring from collapsing under its self-gravity.

\begin{itemize}
  \item Infinite NUT-charge limit
\end{itemize}

It is known that when the NUT charge tends to infinity, the compact direction blows up, and Taub-NUT space becomes four-dimensional flat space. Hence, if we take the infinite NUT-charge limit of the solution (\ref{ER_TN_metric}), we should get a black ring in an asymptotically flat space-time. Indeed, by redefining the coordinates as
\be
t= \sqrt{4n}\,\tilde{t}\,, \qquad \psi= 2n(\tilde{\psi}+\tilde{\phi})\,, \qquad \phi= -\tilde{\psi}+\tilde{\phi}\,,
\ee
and taking out an overall factor $4n$, the metric (\ref{ER_TN_metric}) reduces to the Emparan--Reall black ring \cite{Emparan:2004wy} in the limit $a\rightarrow \infty$, as expected.

\begin{itemize}
  \item Zero NUT-charge limit
\end{itemize}

When the NUT charge $n$ of the solution becomes zero, $\ell_1$, $\ell_3$ and $\ell_4$ become equal to $\frac{\partial}{\partial \phi}$, so the cross terms mixing $\psi$ and $\phi$ in the above metric vanish. This occurs when $a\rightarrow1$. In this limit, we find that the metric $(\ref{ER_TN_metric})$ reduces to the boosted Schwarzschild black string wrapped around a compact dimension (if we recall the identifications (\ref{identifications_coor}), this compact dimension is actually infinitely small). More precisely, if we perform the following parameter redefinitions and coordinate transformations:
\be
m=c\varkappa^2,\quad \sinh\sigma=-\sqrt{\frac{b-c}{c(1-b)}}\,,\quad r=\frac{2\varkappa^2(1+cx)}{x-y}\,,\quad \cos\theta=\frac{1-xy}{x-y}\,,\quad \psi=z\,,
\ee
the limiting metric becomes
\ba
\label{metric_boosted_Sch}
\dif s^2&=&-\left( 1-\frac{2m\cosh^2\sigma }{r}\right) \dif t^2+
\frac{2m\sinh2\sigma }{r} \,\dif t\, \dif z + \left( 1+\frac{2m
\sinh^2\sigma }{r}\right)\dif z^2\nonumber \\
&&  +\left( 1-\frac{2m }{r}\right)^{-1}{\dif r^2}+r^2(\dif\theta^2+\sin^2\theta\, \dif \phi^2)\,,
\ea
which is just the metric of a Schwarzschild black string boosted in the $z$-direction with boost parameter $\sigma$. This is consistent with the fact that the solution (\ref{ER_TN_metric}) does not possess an $S^2$ rotation, which will be clear in the subsequent discussion.

\begin{itemize}
  \item Infinite ring-radius limit
\end{itemize}

In this limit, we obtain the metric of a boosted Schwarzschild black string. To see this, set
\be
\label{abc}
a=1+\frac{h}{\varkappa^2}\,,\qquad
c=\frac{m}{\varkappa^2}\,,\qquad
b=\frac{m}{\varkappa^2}\cosh^2\sigma\,,\qquad
\ee
and define the new coordinates $(r,\theta,z)$ by
\be
x=\cos\theta\,,\qquad
y=-\frac{2\varkappa^2}{r}\,,\qquad
\psi=z+h\phi\,.
\ee
If we then take the $\varkappa\rightarrow\infty$ limit of (\ref{ER_TN_metric}), we obtain the metric (\ref{metric_boosted_Sch}) corresponding to a boosted Schwarzschild black string. 
Note that the form of $a$ in (\ref{abc}) has been chosen to ensure that the NUT charge $n$ remains finite in this limit.
If we further impose the balance condition (\ref{regularity_condition}) in this limit, we recover a black string with boost parameter satisfying $\sinh \sigma=-1$, in agreement with results obtained by Camps et al.~\cite{Camps:2008} using perturbative methods.

\begin{itemize}
  \item Zero ring-radius limit
\end{itemize}

The zero-radius limit of a black ring on Taub-NUT will give a black hole on Taub-NUT. Now, black holes on Taub-NUT have previously been studied in \cite{Ishihara:2005,Wang:2006}, and it is known that they are equivalent to the Kaluza--Klein black holes of \cite{Gibbons:1985,Rasheed:1995}.
In the present case, we obtain a singly rotating Myers--Perry black hole on Taub-NUT (with no rotation along $\ell_1$). This is actually equivalent to a rotating, dyonic Kaluza--Klein black hole \cite{Rasheed:1995,Larsen:1999}, whose angular momentum $J$, electric charge $Q$ and magnetic charge $P$ satisfy $J=PQ$. To obtain the Kaluza--Klein black hole in the form of \cite{Larsen:1999}, we first rewrite $b=(-1+d+2c)/(1+d)$ for some $d\geq1$, and then define the new parameters
\be
m=\sqrt{\frac{(1+a)(1+d)}{2(a+d)}} \,{\varkappa}^{2},\qquad
p=(1+a)\varkappa^2,\qquad
q=(1+d)\varkappa^2.
\ee
The black hole of \cite{Larsen:1999} with $J=PQ$ is then obtained from (\ref{ER_TN_metric}) if we introduce new coordinates $(r,\theta)$ defined by\footnote{Note that there is a typo in (36) of \cite{Larsen:1999}: the sign of $\dif \phi$ term should be a $+$ instead. Also note that in our conventions, Larsen's solution has negative magnetic charge $P$, so that for his solution, the $J=PQ$ limit corresponds to setting his $a$ to be equal to $-\frac{m\sqrt{(p^2-4m^2)(q^2-4m^2)}}{pq+4m^2}$.}
\be
x=-1+\frac{\varkappa^2(1-c)(1+\cos\theta)}{r-m-\varkappa^2\cos\theta}\,, \qquad
y=-1-\frac{\varkappa^2(1-c)(1-\cos\theta)}{r-m-\varkappa^2\cos\theta}\,,
\ee
and take the limit $c\rightarrow1$.

\begin{itemize}
  \item Extremal limit
\end{itemize}

The extremal limit of the black ring on Taub-NUT is taken as $c\rightarrow0$. It turns out that it coincides with the extremal limit of the singly rotating black ring on Taub-NUT found in \cite{Camps:2008}. To map the $c\rightarrow0$ limit of (\ref{ER_TN_metric}) to the form of the extremal solution in \cite{Camps:2008}, we need to rewrite
\be
a=\frac{{\cal C}_2}{{\cal C}_2-{\cal C}_1}\,,\qquad
b=\frac{\lambda{\cal C}_2}{{\cal C}_1-\lambda({\cal C}_2-{\cal C}_1)}\,,\qquad
\varkappa=\sqrt{\frac{\ell \hat R^2}{2}}\,,
\ee
and change coordinates
\be
\psi=-\xi^1,\qquad \phi=-\phi_-\,.
\ee
We refer the reader to \cite{Camps:2008} for a detailed study of this solution. We only note here that in this limit, the black ring actually has a singular horizon of vanishing area, and that the space-time necessarily contains a conical singularity. This is to be contrasted with the extremal doubly rotating black rings to be presented in Sec.~4, which have degenerate but regular horizons of non-vanishing area, and whose space-times do not contain conical singularities.

\newsubsection{Some physical properties}

By introducing the coordinate transformations $\{x=-1+\frac{2\varkappa^2}{r}(1-c)\cos^2\frac{\theta}{2},y=-1-\frac{2\varkappa^2}{r}(1-c)\sin^2\frac{\theta}{2}\}$, the solution (\ref{ER_TN_metric}) approaches
\be
\label{asymtotic_behavior}
\dif s^2\rightarrow -\dif t^2+\left(\dif \psi+2n\cos\theta\,\dif \phi\right)^2+\dif r^2+r^2\left(\dif \theta^2+\sin^2\theta\,\dif \phi^2\right),
\ee
at infinity $r\rightarrow \infty$.
This is the asymptotic geometry of Taub-NUT space with a flat time dimension, comprising of a non-trivial finite $S^1$ fibre bundle over Minkowski space-time $M^{1,3}$. More specifically, surfaces with constant $t$ and $r$ in (\ref{asymtotic_behavior}) are distorted 3-spheres, and they are fibred by the finite circles generated by $\frac{\partial}{\partial\psi}$, over the base space of an $S^2$ parameterised by $\theta$ and $\phi$. This is the well-known Hopf fibration.

It can be shown that for the ranges given by (\ref{range_ER_TN}), the solution (\ref{ER_TN_metric}) is free of naked singularities outside the event horizon. Furthermore, no closed time-like curves (CTCs) have been found in this region despite an extensive numerical search. So if we further impose the condition (\ref{regularity_condition}) to remove the conical singularity along Rod 3, the black ring space-time described by (\ref{ER_TN_metric}) is completely regular outside the event horizon.

Now we calculate the Komar mass and angular momenta of the black ring (\ref{ER_TN_metric}), defined as
\ba
M&=&-\frac{3}{32\pi G_5}\int_{S^3} \epsilon_{abcde}\nabla^d \xi^e,\cr
J_{i}&=&\frac{1}{16\pi G_5}\int_{S^3} \epsilon_{abcde}\nabla^d \psi_i^e,
\ea
where $G_5$ is the gravitational constant in five dimensions. Here $\xi=\frac{\partial}{\partial t}$, $\psi_i$ refers to either $\frac{\partial}{\partial \psi}$ or $\frac{\partial}{\partial \phi}$, $\epsilon$ is the natural volume form of the five-dimensional space-time considered, and the integration is taken over the distorted $S^3$  boundary at infinity of a constant-time hypersurface. Notice that the Killing vector fields $\frac{\partial}{\partial \psi}$ and $\frac{\partial}{\partial \phi}$ themselves do not possess any string or brane-like fixed points---or axes as more commonly referred to---in the space-time; in fact the only fixed point for $\frac{\partial}{\partial \psi}$ or $\frac{\partial}{\partial \phi}$ is the nut point represented by $z_3$ in the rod structure (Fig.~\ref{Figure_ER_TN}). Moreover, the coordinate $\psi$ is even not dimensionless as an angle variable should be. But yet we still formally call the Komar quantities associated to $\frac{\partial}{\partial \psi}$ and $\frac{\partial}{\partial \phi}$ the angular momenta of the system. In fact, $J_{\psi}$ calculated below is more a linear momentum along the compact dimension rather than an angular momentum. Similar remarks have been made for the angular velocities $\Omega_{\psi}$ and $\Omega_{\phi}$ in \cite{Chen:2010zu}. The Komar quantities are calculated to be
\ba
\label{komar_quantity_ER_TN}
M&=&{\frac {6{\varkappa}^{2}\pi nb ( 1-c ) }{ G_5( 1-b )
}}\,,\cr
J_{\psi}&=&{\frac {4{\varkappa}^{
2} \pi  n( 1-c ) \sqrt {b ( 1+b )  ( b-c ) }}{G_5( 1-b )\sqrt {\Psi}
}}\,,\cr
J_{\phi}&=&-{\frac {8{\varkappa}^
{2} \pi n^{2}( 1-c ) \sqrt {b ( 1+b )  ( b-c ) }}{ G_5 ( 1-b
 )\sqrt {\Psi}}}\,.
\ea
For the parameter ranges (\ref{range_ER_TN}), we have $M\geq 0$, $J_{\psi}\geq 0$ and $J_{\phi}\leq 0$. 

A remarkable property of this black ring solution is that it does not possess any rotation along the $S^2$ axis, i.e., the Komar angular momentum associated to $\ell_1=2n\frac{\partial}{\partial \psi}+\frac{\partial}{\partial \phi}$ vanishes:
\be
\label{no_S2_rotation}
J_{S^2}=2nJ_{\psi}+J_{\phi}=0\,.
\ee
This is in contrast to the singly rotating black ring solution found in \cite{Camps:2008}, which has a non-vanishing $J_{S^2}$ in general. The analogous solution in five-dimensional asymptotically flat space-time is the Emparan--Reall black ring, which also does not possess an $S^2$ rotation. This qualifies the solution (\ref{ER_TN_metric}) as the natural generalisation of the Emparan--Reall black ring to Taub-NUT space. And indeed, as we have shown above, the Emparan--Reall black ring is precisely recovered when the Taub-NUT space is blown up to four-dimensional flat space in the limit $n\rightarrow \infty$. 

For the quantities (\ref{komar_quantity_ER_TN}), the following Smarr-like relation can be easily verified:
\be
\label{Smarr_relation}
\frac{2}{3}\,M=\frac{\kappa A}{8\pi G_5}+\Omega_{\psi}J_{\psi}+\Omega_{\phi}J_{\phi}\,,
\ee
where $A$ is the area of the black-hole event horizon, given by
\be
A={\frac {256{\varkappa}^{2}{\pi }^{2} n^{2}c( 1-c )\sqrt {bc ( 1+b )  ( a-c ) } }{ ( a-1)  ( 1-b
 ) ( 1+c ) }}\,.
\ee
We remark that the Komar mass does not take into account the contribution from the Taub-NUT background space-time, so it vanishes in the background limit discussed in the previous subsection.

\newsubsection{Interpretation in Kaluza--Klein theory}

The solution has an interesting interpretation in Kaluza--Klein theory. As we have already mentioned, the Killing vector field $\frac{\partial}{\partial \psi}$ generates finite circles at infinity. We thus perform Kaluza--Klein reduction along this direction. Upon dimensional reduction, the nut of the space-time---represented by the third turning point $z_3$ in the rod structure in Fig.~\ref{Figure_ER_TN}---reduces to the well-known Gross--Perry--Sorkin monopole \cite{Gross:1983,Sorkin:1983}, while the black ring reduces to an electrically charged black hole in four dimensions. The surface gravity $\kappa_{\rm KK}$ and the angular velocity $\Omega_{\rm KK}$ of the black-hole event horizon are calculated to be
\be
\kappa_{\rm KK}=\kappa,\qquad \Omega_{\rm KK}=\Omega_{\phi}\,,
\ee
where the $\kappa$ and $\Omega_{\phi}$ are given in (\ref{ER_surface_gravity}). The monopole and black hole are in general separated by a conical singularity between them. The conical excess angle along the axis $\frac{\partial}{\partial \phi}$ in the reduced metric is
\be
\Delta\phi=2\pi(\kappa_\mathrm{E}-1)\,,
\ee
where $\kappa_\mathrm{E}$ is again given by (\ref{kappa_E}). When (\ref{regularity_condition}) is imposed however, the conical singularity disappears and the system in four dimensions is exactly balanced. There is thus a one-to-one correspondence between balanced configurations in five and four dimensions.

The reduced four-dimensional solution in Kaluza--Klein theory is asymptotically flat. The mass $M_{\rm KK}$, angular momentum $J_{\rm KK}$, electric charge $Q_{\rm KK}$ and magnetic charge $P_{\rm KK}$ of the system are
\ba
M_{\rm KK}&=&\frac{\varkappa^2[\Psi-(1-b)(1-3c)]}{4(1-b)}\,,\quad\hskip0.5cm
J_{\rm KK}=-\frac{\varkappa^4(1-c)(a-1)(a+c)\sqrt{b(1+b)(b-c)}}{2\Psi\sqrt{(1-b)(a-c)}}\,,\cr
Q_{\rm KK}&=&\frac{\varkappa^2(1-c)\sqrt{b(1+b)(b-c)}}{\sqrt{\Psi} (1-b)}\,,\quad
P_{\rm KK}=-n\,,
\ea
where the gravitational constant in four dimensions $G_4$ has been set to unity. It can be verified that in the background limit $b,c\rightarrow 0$, the well-known result $M_{\rm KK}=\frac{n_0}{2}$ is recovered. If we take into consideration the fact that $G_5=l_{\rm c}G_4$, where $l_{\rm c}=8\pi n$ is the length of the compact dimension at infinity, we then get the relations
\be
\label{JKK_QKK}
J_{\rm KK}=J_{\phi}\,,\qquad Q_{\rm KK}=2J_{\psi}\,.
\ee
These results essentially say that upon Kaluza--Klein reduction, angular momentum along the compact direction in five dimensions reduces to electric charge in four dimensions, while angular momentum along $\frac{\partial}{\partial \phi}$ becomes the usual angular momentum in four dimensions. The area of the event horizon of the black hole is simply $A_{\rm KK}=A/l_{\rm c}$. We point out that the magnetic charge is entirely carried by the monopole, while the electric charge is entirely carried by the black hole. Hence in Kaluza--Klein theory, we have a purely electrically charged black hole in balance with a purely magnetically charged monopole, for appropriately chosen parameters. Moreover, the black hole is static, in the sense that if the monopole is removed from the space-time (by setting its charge to zero), we recover a space-time with vanishing angular momentum.

Although both the monopole and black hole are individually static objects, the space-time possesses a non-zero angular momentum in general. Note that Eq.~(\ref{no_S2_rotation}) reduces to the relation
\be
J_{\rm KK}=P_{\rm KK}Q_{\rm KK}\,,
\ee
which shows that the angular momentum arises solely from the fact that we have a magnetically charged object in superposition with an electrically charged object. It is interesting to note that this formula holds regardless of the distance between the two objects. Indeed, a similar phenomenon is known to occur in classical electrodynamics \cite{Jackson}.

\newsection{Pomeransky--Sen'kov black ring on Taub-NUT}

The Pomeransky--Sen'kov (PS) black ring on Taub-NUT is obtained from the above-mentioned six-parameter solution by setting
\ba
\label{balance_condition_PS_TN}
b&=&{\frac {2c \left( a-{c}^{2} \right) }{   \left( 1
+{c}^{2} \right)\left( a+c \right) -2c \left( 1+c \right) }}\,,\cr
C_3&=&-\frac{2\varkappa^4c^3\left[a(1-c)-c(1+3c)\right]\sqrt{2(a-c)}}{(1+c)\sqrt{(1-c^2)(a+c)(a-c^2)}}\,C_2\,.
\ea
The value for $b$ is just that in (\ref{regularity_condition}), which was imposed to eliminate the conical singularity in the Emparan--Reall black ring on Taub-NUT. These two conditions ensure that the direction of the left semi-infinite rod is identical to that of the inner axis of the solution. Again, we need to perform a linear transformation on the $G$-matrix and appropriately choose the integration constant, to eliminate the apparent Dirac--Misner singularity and to bring the directions of all the rods to the correct orientation. The final solution has four parameters: $\varkappa$, $a$, $c$, and $C_2$. It is not difficult to calculate the rod structure of this solution, which is qualitatively the same as Fig.~\ref{Figure_ER_TN} (now with $\kappa_{\rm E}=1$). The parameter $a$, as before, characterises the NUT charge; when it is taken to infinity, we can recover (after taking out an overall factor) the PS black ring. The parameter $C_2$, roughly speaking, characterises the $S^2$ rotation; if it is set to zero, we recover the (regular) Emparan--Reall black ring on Taub-NUT discussed in the previous section. These facts suggest that we have indeed obtained the correct generalisation of the PS black ring to Taub-NUT space.

Unfortunately, this solution is complicated enough that we are not going to present its full explicit form. In what follows, we shall restrict ourselves to the extremal limit. In this configuration, the $S^2$ rotation is saturated, so that the surface gravity on the event horizon becomes zero. Depending on the sign of $C_2$, and so the sign of the $S^2$ rotation, there are two distinct classes of solutions that we are going to discuss separately.

\newsubsection{Class I extremal PS black ring on Taub-NUT}

To obtain the Class I extremal PS black ring on Taub-NUT, we start from the four-parameter solution after imposing (\ref{balance_condition_PS_TN}), set
\be
C_2=-\frac{1}{2\varkappa^2 c}\,[c^{-1}+2a^{-1}-(\alpha-2)]\,,
\ee
and then take the limit $c\rightarrow 0$. One finds that although $C_2$ blows up in this limit, all components of the direction of, say, the first rod remain finite, and so do the metric components. We then obtain the solution:
\ba
\label{metric_PS_ex_TN_classI}
\dif s^2&=&-\frac{K(x,y)}{H(x,y)}\,(\dif t+\omega_1\,\dif \psi+\omega_2\,\dif \phi)^2+\frac{F(x,y)}{K(x,y)}\,(\dif \psi+\omega_3\, \dif \phi)^2-\frac{4\varkappa^4G(x)G(y)H(x,y)}{(x-y)^4F(x,y)}\,\dif \phi^2\cr
&&+\frac{\varkappa^4 H(x,y)}{a(\alpha-1)^2(x-y)^3}\left(\frac{\dif x^2}{G(x)}-\frac{\dif y^2}{G(y)}\right),
\ea
where
\ba
\omega_1&=&\frac{2\sqrt{2a}(x-y)J(x,y)}{K(x,y)}\,,\cr
\omega_2&=&\frac{2\sqrt{2a}\varkappa^2}{(\alpha-1)(x-y)K(x,y)}\Big\{(x-y)K(x,y)-(a\alpha-a-\alpha-1)(x-y)^2J(x,y)\cr
&&-2(\alpha-1)\big [a(a-1)(x-y)^2(2\alpha-2x-yG(x))+a(x+y)[G(x)G(y)-2(\alpha-1)(x-y)]\cr
&&-(a-1)(1+x)G(x)(x-xy-3y+1+2\alpha)\big]\Big\}\,,\cr
\omega_3&=&-\frac{\varkappa^2}{(\alpha-1)(x-y)F(x,y)}\Big\{a(a-1)(\alpha-1)(x^2+y^2-2)[G(x)G(y)+4(\alpha^2-1)]\cr
&&+2G(x)G(y)[4a(\alpha-1)(1-a+y)+a(x^2-y^2)+2(1+\alpha-2a)(x-y)]\Big\}\,,
\ea
and the functions $G$, $F$, $H$, $J$ and $K$ are defined as
\ba
G(x)&=&1-x^2,\cr
F(x,y)&=&[4(a-1)-a(x+y)]G(x)G(y)-4a(\alpha-1)^2(x+y)\,,\cr
H(x,y)&=&4(a-1)(1-x)(1+y)G(x)-4a(x+y)(x^2-x+y+1-2\alpha)\cr
      &&+a(ax-ay-2x)[(2\alpha+x-y)^2+(xy+2x-1)(xy-2x-1)]\,,\cr
J(x,y)&=&-a(x+y)(x-y+2\alpha-2)-(a-1)(1-x)(x+y-xy+1+2\alpha)\,,\cr
K(x,y)&=&4(a-1)(1+x)(1-y)G(x)+4a(x+y)(x^2-x+y-1-2xy+2\alpha)\cr
      &&+a(ax-ay-2x)[(2\alpha-x+y)^2+(xy+2y-1)(xy-2y-1)]\,.
\ea
There are three parameters $\varkappa$, $a$ and $\alpha$ in the above metric, satisfying the following constraints:
\be
\label{range_PS_TN_classI}
a>1\,, \qquad \alpha>1\,,\qquad \varkappa>0\,.
\ee
Similar to the Emparan--Reall black ring on Taub-NUT, $\varkappa$ roughly speaking sets the scale of the solution, while $a$ determines the NUT charge. The third parameter $\alpha$ controls the rotations of the black ring. The coordinates used here are the same as those of (\ref{ER_TN_metric}) but with $c=0$, so now the horizon is located at $y=-\infty$, and the ranges of the coordinates are
\be
\label{coordinate_range_classI_ex_PS_TN}
-\infty<t<\infty\,, \qquad -\infty\leq y\leq -1\leq x\leq 1\,.
\ee

\begin{figure}[t]
\begin{center}
\includegraphics{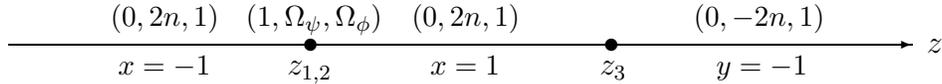}
\caption{The rod structure of the extremal Pomeransky--Sen'kov black ring on Taub-NUT, where $n$ is the NUT-charge parameter. The point $z_{1,2}$ represents the extremal horizon at $y=-\infty$ with zero surface gravity, while the turning point $z_3$ represents the location of the nut.}
\label{Figure_PS_TN}
\end{center}
\end{figure}

We remark that the rod structure formalism can also be applied to the extremal solutions presented here. It seems to be a common feature that an extremal horizon is represented by a point in the rod structure \cite{Figueras:2009ci}. However, these points are not simply the turning points defined in \cite{Chen:2010zu} for a given solution, since the latter are positions on the $z$-axis where the kernel of $G$-matrix is more than one-dimensional. It will be clear in the rod structure of the above solution, that the kernel of the $G$-matrix is either one or two-dimensional at the point representing the extremal horizon, depending on how the point under consideration is approached; in fact this point is more like a rod with vanishing length along the $z$-axis in the rod structure, so it has two endpoints which are the true turning points, and an interior which we can assign a direction. We also point out that for an extremal horizon, its surface gravity is zero, and so its direction in the rod structure cannot be normalised to have unit surface gravity as was done \cite{Chen:2010zu} for the non-extremal case. Roughly speaking, the rod structure of the above solution is like that of (\ref{ER_TN_metric}), with the horizon-rod made doubly rotating and shrunk to a point. The Weyl--Papapetrou coordinates are related to the above coordinates by (\ref{C-metric}) with $c=0$. The rod structure is shown in Fig.~\ref{Figure_PS_TN} and can be summarised as follows ($z_{1,2}\equiv 0$):
\begin{itemize}
\item Rod 1: a semi-infinite space-like rod located at $(\rho=0, z\leq z_1)$ or $(x=-1,-\infty\leq  y< -1)$, with direction $\ell_1=(0,2n,1)$, where the NUT charge $n$ is given by
\be
n={\frac { {\varkappa}^{2} ( a-1 )(\alpha+1) }{2(\alpha-1)}}\,.
\ee

\item Extremal horizon: a point located at $(\rho=0, z=z_{1,2})$ or $(-1\leq x\leq 1,y=-\infty)$, with direction $\ell_2=(1,\Omega_{\psi},\Omega_{\phi})$ and zero surface gravity, where the angular velocities $\Omega_\psi$ and $\Omega_\phi$ are given by
\be
\label{angular_velocity_classI_extremal}
\Omega_{\psi}=-\frac{a(\alpha-1)-(\alpha+1)}{2\sqrt{2a}}\,,\qquad
\Omega_{\phi}=-\frac{\alpha-1}{2\sqrt{2a}\varkappa^2}\,.
\ee

\item Rod 3: a finite space-like rod located at $(\rho=0, z_2\leq z\leq z_3)$ or $(x=1,-\infty\leq y\leq -1)$, with direction $\ell_3=(0,2n,1)$.
\item Rod 4: a semi-infinite space-like rod located at $(\rho=0, z\geq z_3)$ or $(-1<x\leq 1,y=-1)$, with direction $\ell_4=(0,-2n,1)$.
\end{itemize}

Again $\{\ell_1,\ell_4\}$ are identified as the two independent $2\pi$-periodic generators of the $U(1)\times U(1)$ isometry subgroup of the solution. The space-time topology outside the event horizon is the same as that of the Emparan--Reall black ring on Taub-NUT, except that in the present case, the conical singularity has already been removed. In particular, the event horizon has a topology $S^1\times S^2$, and $\ell_1$ generates the rotational symmetry of the $S^2$.

Infinity of the space-time is located at $(x,y)\rightarrow(-1,-1)$, where it approaches Taub-NUT space with a flat time dimension, in the form (\ref{asymtotic_behavior}). The background limit itself can be taken as $\alpha\rightarrow\infty$; in this case we recover the Taub-NUT background space (\ref{SD_TN}) after the transformations (\ref{bg_transf}).

The extremal PS black ring \cite{Pomeransky:2006,Elvang:2007hs} is recovered in the limit when the NUT charge $n$ becomes infinite. More precisely, to take this limit, we first need to perform the following parameter redefinition and coordinate transformations:
\ba
\label{large_NUT_charge_limit_exPStn_classI}
\alpha&=&\frac{1+\mu^2}{2\mu}\,,\qquad  x=\frac{\tilde{x}+\mu}{1+\mu \tilde{x}}\,,\qquad ~~~~y=\frac{\tilde{y}+\mu}{1+\mu \tilde{y}}\,,\cr
t&=& \sqrt{4n}\,\tilde{t}\,, \qquad ~\psi= 2n(\tilde{\psi}+\tilde{\phi})\,, \qquad \phi= -\tilde{\psi}+\tilde{\phi}\,,
\ea
and then set $a\rightarrow \infty$ (after taking out an overall factor $4n$). The resulting metric is identical to the extremal PS black ring in the form presented in Eq.~(B.1) of \cite{Chen:2011jb} (setting $\nu=\mu$), but with an $S^2$ rotation in the \emph{negative} sense. It is also clear that $\alpha$ is a measure of the ratio of the $S^1$ and $S^2$ rotations in this limit.

On the other hand, when the NUT charge $n$ tends to zero, the metric (\ref{metric_PS_ex_TN_classI}) reduces to the boosted extremal Kerr black string with negative $S^2$ rotation. To see this, we need to define
\be
m=\frac{\varkappa^2}{\alpha-1}\,,\qquad r=m+\frac{2\varkappa^2}{x-y}\,,\qquad \cos \theta=\frac{1-xy}{x-y}\,, \qquad \psi=z\,,
\ee
and then in the limit $a\rightarrow 1$, the metric (\ref{metric_PS_ex_TN_classI}) reduces to the following solution:
\ba
\label{boosted_exKerr_string}
\dif s^2&=&-\left( 1-\frac{2m r \cosh^2\sigma }{\Sigma}\right) \dif t^2+
\frac{2m r \sinh2\sigma }{\Sigma} \,\dif t\, \dif z + \left( 1+\frac{2m r
\sinh^2\sigma }{\Sigma}\right)\dif z^2\nonumber \\
 & & +\frac{(r^2+a_{\rm K}^2)^2-\Delta a_{\rm K}^2
\sin^2\theta}{\Sigma}\,\sin^2\theta\, \dif \phi^2
  -\frac{4m r \cosh\sigma }{\Sigma}\,a_{\rm K}\sin^2\theta\, \dif t\, \dif \phi\nonumber \\
&&  -\frac{4m r \sinh\sigma }{\Sigma}\,
  a_{\rm K}\sin^2\theta\, \dif z\, \dif \phi
+\Sigma\left(\frac{\dif r^2}{\Delta}+\dif\theta^2\right),
\ea
where
\be
 \Delta = r^2-2mr+a_{\rm K}^2 \,, \qquad
 \Sigma = r^2+ a_{\rm K}^2\cos^2\theta \,,
\ee
and with the parameters $a_{\rm K}$ and $\sigma$ fixed as
\be
\label{extremal_condition}
a_{\rm K}=-m\,, \qquad\sinh\sigma=-1\,.
\ee
One immediately recognises this solution to be the extremal Kerr black string with negative $S^2$ rotation, boosted along the $z$ direction with a boost parameter $\sigma$.

To take the infinite ring-radius limit of (\ref{metric_PS_ex_TN_classI}), we first need to redefine the coordinates $(x,y,\psi)$ as
\be
\label{large_rad_limit1}
x=\cos \theta,\qquad y=-\frac{2\varkappa^2}{r-m}\,,\qquad \psi=z+h\phi\,,
\ee
and the parameters $(a,\alpha)$ as
\be
\label{large_rad_limit2}
a=1+\frac{h}{\varkappa^2}\,,\qquad\alpha=\frac{\varkappa^2}{{m}}\,,
\ee
and then set $\varkappa\rightarrow \infty$. The resulting metric is identified with the boosted extremal Kerr string with a specific boost parameter, exactly given by (\ref{boosted_exKerr_string}) and (\ref{extremal_condition}).

It can be shown that for the specified ranges of parameters (\ref{range_PS_TN_classI}), the space-time is free of naked singularities outside the event horizon. Also no CTCs have been found in this region despite a numerical search for them.

The Komar quantities can be easily calculated:
\be
M=\frac{12\pi\varkappa^2n}{G_5(\alpha-1)}\,,\quad
J_{\psi}=\frac{4\sqrt{2}\pi\varkappa^2n}{G_5\sqrt{a}(\alpha-1)}\,,\quad
J_{\phi}=-\frac{4\sqrt{2}\pi\varkappa^4n(a\alpha-\alpha+3a-1)}{G_5\sqrt{a}(\alpha-1)^2}\,.
\ee
It is straightforward to show that this black ring carries negative $S^2$ rotation:
\be
\label{S2_PS_TN_classI}
J_{S^2}=-\frac{8\pi\sqrt{2a}\varkappa^4n}{G_5(\alpha-1)^2}<0\,.
\ee
This is to be compared with the Class II extremal black ring on Taub-NUT to be discussed in the next subsection, which has positive $S^2$ rotation. The event horizon has zero surface gravity $\kappa=0$ and a finite area
\be
A=\frac{64\pi^2\sqrt{2a}\varkappa^4n}{(\alpha-1)^2}\,.
\ee
With the angular velocities of the horizon given by (\ref{angular_velocity_classI_extremal}), the Smarr-like relation (\ref{Smarr_relation}) can then be verified directly. It is easy to check that the entropy $S=\frac{A}{4G_5}$ satisfies the very simple formula
\be
\label{entropy_classI_exPS_TN}
S=-2\pi J_{S^2}\,.
\ee
It was shown in \cite{Hollands:2011sy} that the inequality $S\geq2\pi |J_{S^2}|$ holds quite generally for black rings, so equality is satisfied for the present extremal black ring solution. Such a formula was first observed for the extremal PS black ring by Reall \cite{Reall:2007jv}, and it led him to derive the entropy of that solution from a microscopic counting of states. We expect that a similar microscopic counting of states would enable one to derive the entropy formula of the present solution.

Upon dimensional reduction in Kaluza--Klein theory, the above solution describes a purely electrically charged rotating extremal black hole, in balance with a magnetic monopole in an asymptotically flat space-time. More specifically, the extremal black ring from a five-dimensional perspective reduces to the purely electrically charged extremal black hole, while the nut of the solution reduces to the magnetic monopole in Kaluza--Klein theory. Once again, the rotation exactly balances the electromagnetic interaction between these two objects. The conserved charges of the Kaluza--Klein system are
\ba
M_{\rm KK}&=&\frac{\varkappa^2(a\alpha-\alpha+a+5)}{4(\alpha-1)}\,,\qquad J_{\rm KK}=-\frac{\varkappa^4(a\alpha-\alpha+3a-1)}{\sqrt{2a}(\alpha-1)^2}\,,\cr
Q_{\rm KK}&=&\frac{\sqrt{2}\varkappa^2}{\sqrt{a}(\alpha-1)}\,,\qquad\quad\hskip1.2cm P_{\rm KK}=-n\,.
\ea
It is clear that $M_{\rm KK}>0$, $J_{\rm KK}<0$, $Q_{\rm KK}>0$ and $P_{\rm KK}\leq 0$. Again, if we take into consideration $G_5= l_{\rm c}\equiv 8\pi n$, we get the relations (\ref{JKK_QKK}). In Kaluza--Klein theory, (\ref{S2_PS_TN_classI}) reduces to
\be
J_{\rm KK}<P_{\rm KK}Q_{\rm KK}\,.
\ee
Considering the signs of $P_{\rm KK}$ and $Q_{\rm KK}$, we actually have $|J_{\rm KK}|>|P_{\rm KK}Q_{\rm KK}|$. Contrast this to the formula $|J_{\rm KK}|=|P_{\rm KK}Q_{\rm KK}|$ for the Kaluza--Klein system obtained from the Emparan--Reall black ring on Taub-NUT. In this system, the black hole may be thought of as co-rotating with respect to the angular momentum carried by the electromagnetic fields.

The black hole is extremal, so it has zero surface gravity, but it has a finite horizon-area $A_{\rm KK}=A/l_{\rm c}$. It is interesting to note that the entropy of this black hole is given by the formula
\be
\label{entropy_KK}
S_{\rm KK}=2\pi|J_{\rm KK}-P_{\rm KK}Q_{\rm KK}|\,,
\ee
which follows from the reduction of (\ref{entropy_classI_exPS_TN}).
Its event horizon has angular velocity $\Omega_{\rm KK}=\Omega_{\phi}$. When the monopole is removed, the black hole lies on surface {\bf S} (in the $P=0$ plane) in the space of extremal black-hole solutions in Kaluza--Klein theory as described by Rasheed (Fig.\ 2 of \cite{Rasheed:1995}).
The monopole itself lies in the intersection between surfaces {\bf S} and {\bf W}, on the $P$ axis.

We note an interesting relation between the masses from the four and five-dimensional perspectives ($G_5=l_{\rm c}$):
\be
\label{total_mass_formula}
M_{\rm KK}=M+\frac{n}{2}\,.
\ee
This formula suggests that the total mass of the Kaluza--Klein system comes from the black hole mass and the monopole mass. It does not hold in the non-extremal case in general.

\newsubsection{Class II extremal PS black ring on Taub-NUT}

To obtain the Class II extremal PS black ring on Taub-NUT, we start with the four-parameter solution, set
\be
C_2=\frac{1}{2\varkappa^2 c}\,[c^{-1}-(\alpha-2)]\,,
\ee
and then take the limit $c\rightarrow 0$. We then perform a linear transformation on the $G$-matrix and make a suitable choice of the integration constant, to bring the solution into a form as close to (\ref{metric_PS_ex_TN_classI}) as possible:
\ba
\label{metric_PS_ex_TN_classII}
\dif s^2&=&-\frac{K(x,y)}{H(x,y)}\,(\dif \hat{t}+\omega_1\,\dif \hat{\psi}+\omega_2\,\dif \phi)^2+\frac{F(x,y)}{K(x,y)}\,(\dif \hat{\psi}+\omega_3\, \dif \phi)^2-\frac{4\varkappa^4G(x)G(y)H(x,y)}{(x-y)^4F(x,y)}\,\dif \phi^2\cr
&&+\frac{\varkappa^4 H(x,y)}{a(\alpha+1)^2(x-y)^3}\left(\frac{\dif x^2}{G(x)}-\frac{\dif y^2}{G(y)}\right),
\ea
where
\ba
\omega_1&=&\frac{2\sqrt{2a}(x-y)J(x,y)}{K(x,y)}\,,\cr
\omega_2&=&\frac{2\sqrt{2a}\varkappa^2}{a(\alpha+1)(x-y)K(x,y)}\Big\{-(x-y)K(x,y)-a(a\alpha+a-\alpha+1)(x-y)^2J(x,y)\cr
&&+2a(\alpha+1)\big [a(a-1)(x-y)^2(2\alpha-2x-yG(x))+a(x+y)[G(x)G(y)-2(\alpha+1)(x-y)]\cr
&&-(a-1)(1-x)G(x)(xy+x-3y-1+2\alpha)\big]\Big\}\,,\cr
\omega_3&=&-\frac{\varkappa^2}{(\alpha+1)(x-y)F(x,y)}\Big\{a(a-1)(\alpha+1)(x^2+y^2-2)[G(x)G(y)+4(\alpha^2-1)]\cr
&&+2G(x)G(y)[4a(\alpha+1)(1-a-y)-a(x^2-y^2)+2(1-\alpha-2a)(x-y)]\Big\}\,,
\ea
and the functions $G$, $F$, $H$, $J$ and $K$ are defined as
\ba
G(x)&=&1-x^2,\cr
F(x,y)&=&-[4(a-1)+a(x+y)]G(x)G(y)-4a(\alpha+1)^2(x+y)\,,\cr
H(x,y)&=&-4(a-1)(1+x)(1-y)G(x)-4a(x+y)(x^2+x-y+1+2\alpha)\cr
      &&+a(ax-ay-2x)[(x-y+2\alpha)^2+(xy+2x-1)(xy-2x-1)]\,,\cr
J(x,y)&=&-a(x+y)(x-y+2\alpha+2)+(a-1)(1+x)(xy+x+y-1+2\alpha)\,,\cr
K(x,y)&=&-4(a-1)(1-x)(1+y)G(x)+4a(x+y)(x^2-2xy+x-y-1-2\alpha)\cr
      &&+a(ax-ay-2x)[(x-y-2\alpha)^2+(xy+2y-1)(xy-2y-1)]\,.
\ea
However, in the present coordinates, the rod structure of the solution is not in the correct orientation, so asymptotically this solution does not possess the form (\ref{asymtotic_behavior}). To bring it to this asymptotic form, we need to perform the boost:\footnote{The boost (\ref{boost_classII}) can be absorbed into the linear transformation on the $G$-matrix we mentioned just above Eq.~(\ref{metric_PS_ex_TN_classII}), and the solution (\ref{metric_PS_ex_TN_classII}) can then be directly written in the coordinates $(t,\psi,\phi,x,y)$. However, this will make it more complicated in form.}
\be
\label{boost_classII}
\hat{t}=t\cosh \lambda -\psi\sinh \lambda\,,\qquad \hat{\psi}=-t\sinh \lambda +\psi\cosh \lambda\,,
\ee
with
\be
\sinh \lambda=\sqrt{\frac{8}{\Phi}}\,,\qquad \Phi\equiv a(\alpha-1)^2-8\,.
\ee

The above solution has three parameters $\varkappa$, $a$ and $\alpha$, satisfying the following constraints:
\be
\label{range_PS_TN_classII}
a>1\,, \qquad \alpha>1+\sqrt{\frac{8}{a}}\,,\qquad \varkappa>0\,.
\ee
Note that the lower bound of $\alpha$ corresponds to $\Phi=0$. The interpretations of the three parameters are similar to those of the Class I solution. The horizon is located at $y=-\infty$, and the ranges of the coordinates $(t,x,y)$ are the same as (\ref{coordinate_range_classI_ex_PS_TN}).

{}From now on, we analyse this solution in the coordinates $(t,\psi,\phi,x,y)$. Its rod structure is easily calculated, and it is the same as that of the Class I solution (Fig.~\ref{Figure_PS_TN}), but with different values of the NUT charge $n$ and angular velocities $\Omega_{\psi}$ and $\Omega_{\phi}$:
\be
\label{NUT_charge_classII}
n=\frac{\varkappa^2(a-1)\sqrt{\Phi}}{2(\alpha+1)\sqrt{a}}\,,\qquad \Omega_{\psi}=\frac{\sqrt{2}[a^2(\alpha^2-1)-\Phi]}{4a^{3/2}(\alpha+1)}\,,\qquad \Omega_{\phi}=\frac{\sqrt{2\Phi}}{4a\varkappa^2}\,.
\ee
In particular, the horizon located at $y=-\infty$ is extremal and has zero surface gravity. As with the Class I solution, $\{\ell_1,\ell_4\}$ are identified as the two independent $2\pi$-periodic generators of the $U(1)\times U(1)$ isometry subgroup of this solution.

Infinity of the space-time is located at $(x,y)\rightarrow(-1,-1)$, where it approaches Taub-NUT space with a flat time dimension, in the form (\ref{asymtotic_behavior}). The background limit itself can be taken as $\alpha\rightarrow\infty$, just as in the Class I solution.

{}From this solution we can also recover the extremal PS black ring \cite{Pomeransky:2006,Elvang:2007hs} in the limit when the NUT charge $n$ becomes infinite. This limiting procedure is the same as that for the Class I solution, except that the NUT charge now takes the new value in (\ref{NUT_charge_classII}). Then in this limit we recover the extremal PS black ring with an $S^2$ rotation in the \emph{positive} sense.

On the other hand, when the NUT charge $n$ tends to zero, the metric (\ref{metric_PS_ex_TN_classII}) reduces to the boosted extremal Kerr black string with positive $S^2$ rotation. To see this, we need to define
\be
m=\frac{\varkappa^2}{\alpha+1}\,,\qquad r=m+\frac{2\varkappa^2}{x-y}\,,\qquad \cos \theta=\frac{1-xy}{x-y}\,, \qquad \psi=z\,,
\ee
and then in the limit $a\rightarrow 1$ the metric (\ref{metric_PS_ex_TN_classII}) (in the coordinates $(t,\psi,\phi,x,y)$) reduces to the solution (\ref{boosted_exKerr_string}), now with the parameters
\be
a_{\rm K}=m\,,\qquad \sinh \sigma=-\frac{\alpha+3}{\sqrt{(\alpha-1)^2-8}}\,.
\ee

To take the infinite ring-radius limit of (\ref{metric_PS_ex_TN_classII}), we first 
perform the same redefinitions as in (\ref{large_rad_limit1}) and (\ref{large_rad_limit2}), and then set $\varkappa\rightarrow \infty$. The resulting metric is identified with the boosted extremal Kerr string, given by (\ref{boosted_exKerr_string}) with the parameters specified as
\be
a_{\rm K}=m\,, \qquad\sinh\sigma=-1\,.
\ee

For the specified ranges of parameters (\ref{range_PS_TN_classII}), the space-time is free of naked singularities outside the event horizon. Also no CTCs have been found in this region despite a numerical search for them.

The Komar quantities of the solution are calculated to be
\ba
M&=&\frac{12\pi\varkappa^2n(a\alpha-a+2)}{G_5\Phi}\,,\cr 
J_{\psi}&=&\frac{4\sqrt{2a}\pi\varkappa^2n(\alpha+3)}{G_5\Phi}\,,\cr
J_{\phi}&=&-\frac{4\sqrt{2}\pi\varkappa^4n(a\alpha-\alpha+a-3)}{G_5\sqrt{\Phi}(\alpha+1)}\,.
\ea
Then it is easy to show that, as mentioned in the previous subsection, this black ring carries positive $S^2$ rotation:
\be
\label{S2_PS_TN_classII}
J_{S^2}=\frac{8\sqrt{2}\pi\varkappa^4an}{G_5\sqrt{\Phi}(\alpha+1)}>0\,.
\ee
Hence, it is the sign of the $S^2$ rotation that distinguishes the two classes of extremal PS black rings on Taub-NUT. When the NUT charge is finite, positive and negative $S^2$ rotations result in physically different configurations (that are not isometric to each other). These two classes, however, become isometric to each other in the infinite NUT-charge limit, when both of them reduce to the extremal PS black ring. We also note that the two classes interpolate in the background limit $\alpha\rightarrow\infty$, for which $J_{S^2}=0$. This is consistent with the results of Sec.~3, that there is no regular extremal black ring on Taub-NUT when $J_{S^2}=0$.

The event horizon has zero surface gravity $\kappa=0$ and a finite area
\be
A=\frac{64\sqrt{2}\pi^2\varkappa^4an}{\sqrt{\Phi}(\alpha+1)}\,.
\ee
The Smarr-like relation for the black ring can then be verified directly. Its entropy $S$ then satisfies
\be
\label{entropy_classII_exPS_TN}
S=2\pi J_{S^2}\,.
\ee
This formula is a reflection of the fact that the black ring is extremal, and a microscopic counting of states might be able to explain it.

Like the Class I solution, upon dimensional reduction in Kaluza--Klein theory, the above solution describes a purely electrically charged rotating extremal black hole, in exact balance with a magnetic monopole in an asymptotically flat space-time. The conserved charges of the Kaluza--Klein system are
\ba
M_{\rm KK}&=&\frac{\varkappa^2(\alpha-1)[\Phi(a-1)+6a\alpha+2a+8]}{4\Phi(\alpha+1)}\,,\qquad
J_{\rm KK}=-\frac{\varkappa^4(a\alpha-\alpha+a-3)}{\sqrt{2\Phi}(\alpha+1)}\,,\cr
Q_{\rm KK}&=&\frac{\varkappa^2\sqrt{2a}(\alpha+3)}{\Phi}\,,\qquad\hskip3.85cm
P_{\rm KK}=-n\,.
\ea
It can be checked that by setting $G_5=l_{\rm c}$, we get the relations (\ref{JKK_QKK}), but the mass formula (\ref{total_mass_formula}) does not hold anymore. It might be worthwhile to investigate why the formula holds for the previous class of solutions, but breaks down for this class.

In Kaluza--Klein theory, the relation (\ref{S2_PS_TN_classII}) reduces to
\be
\label{inequality_classII}
J_{\rm KK}>P_{\rm KK}Q_{\rm KK}\,.
\ee
Considering the signs of the quantities $P_{\rm KK}\leq 0$ and $Q_{\rm KK}>0$, we actually have $J_{\rm KK}>-|P_{\rm KK}Q_{\rm KK}|$. Note however that, unlike the Class I solution, the sign of $J_{\rm KK}$ in the Class II solution is indefinite: it can be either negative, zero, or positive. To see this, we consider the infinite and zero NUT-charge limits. For the former, $a\rightarrow \infty$, and we have $J_{\rm KK}<0$; while for the latter, we have shown it corresponds to the boosted extremal Kerr black string with positive $S^2$ rotation, and we have $J_{\rm KK}>0$. For some values of parameters, we can then have $J_{\rm KK}=0$, so when measured at infinity, the whole system does not carry any angular momentum. Physically this may be interpreted as the extremal black hole counter-rotating with respect to the angular momentum carried by the electromagnetic field of the system, so that together they give rise to a vanishing total angular momentum.

The black hole is again extremal with zero surface gravity and a finite horizon-area $A_{\rm KK}=A/l_{\rm c}$. Its entropy $S_{\rm KK}$ is given by the same expression (\ref{entropy_KK}) as in the Class I solution. Its event horizon has angular velocity $\Omega_{\rm KK}=\Omega_{\phi}$. When the monopole is removed, the black hole lies on Rasheed's surface {\bf S} \cite{Rasheed:1995}, in the $P=0$ plane.

\newsection{Discussion}

In this paper, we have constructed several new vacuum solutions describing rotating black rings on Taub-NUT, improving on and extending previously known solutions in the literature. The first is an $S^1$-rotating black ring on Taub-NUT, which can be regarded as the natural generalisation of the Emparan--Reall black ring on flat space. The second set of solutions---which actually consists of two separate classes---describes an extremal doubly rotating black ring on Taub-NUT. This set of solutions is free of conical singularities, and can be regarded as the natural generalisation of the extremal Pomeransky--Sen'kov black ring on flat space. All these solutions admit a Kaluza--Klein reduction to four dimensions, in which they describe an electrically charged black hole a finite distance away from a magnetic monopole.

There are a number of possible generalisations and applications of the solutions obtained in this paper. For instance, it is straightforward to introduce an extra black hole at the nut, thus obtaining what might be called a black Saturn on Taub-NUT. Like the black Saturn on flat space \cite{Elvang:2007rd}, it should be possible to tune the $S^1$ rotation of the black ring to eliminate the conical singularity in the space-time. Such a solution, when reduced to four dimensions, would describe an electric and a magnetic black hole in equilibrium a finite distance from each other. The crossed electric and magnetic fields will endow the space-time with an angular momentum, in addition to the intrinsic angular momenta of the black holes themselves. The simplest and perhaps most interesting subclass of this system consists of an $S^1$-rotating black ring surrounding a static black hole at the nut. When reduced to four dimensions, it would describe a static electric black hole in equilibrium with a static magnetic black hole. A study of this solution is in progress.

As mentioned in the introduction, the solutions we have obtained can be embedded in Type IIA string theory, in which case they will describe non-supersymmetric configurations of D0- and D6-branes. In particular, the extremal PS on Taub-NUT solutions will give balanced configurations of zero-temperature D0- and D6-branes. It would be interesting to study the physics of these D-brane configurations following the methods of \cite{Camps:2008}. We remark that non-supersymmetric black rings on Taub-NUT have also been considered within a string theory context by Bena et al.~\cite{Bena:2009ev,Bena:2011}, although their solutions are inequivalent to the ones obtained in this paper.

A longer-term goal would be to move away from a Taub-NUT background, and construct rotating black hole/ring solutions on other backgrounds, which can essentially be any gravitational instanton \cite{Chen:2010zu}. These will generalise the static black-hole solutions found in \cite{Chen:2010ih}. A very general class of solutions can in fact be obtained by extending the ISM construction in Sec.~2, by removing an additional fifth soliton at $z_4$ with a BZ vector $(0,0,1)$ in step 1, and then adding it back in step 2 with a more general BZ vector $(0,C_4,1)$. It can be shown that, after joining up the first two rods as well as the last two rods, the generated solution describes a doubly rotating black hole on a Euclidean Kerr-NUT background (possibly with Dirac--Misner string singularities present).\footnote{In fact, the static black holes on the Euclidean Kerr and Taub-bolt instantons presented in \cite{Chen:2010ih} were first identified as special cases of the static limit of this solution.} The rotating black rings on Taub-NUT presented in this paper are obviously special cases of this solution. Other regular classes contained in this general solution would include doubly rotating black holes on the Euclidean Kerr and Taub-bolt instantons. We hope to return to these issues in the future.

\bigbreak\bigskip\bigskip\centerline{{\bf Acknowledgements}}
\nobreak\noindent We are grateful to Roberto Emparan for his comments on the manuscript. ET wishes to thank Susan Scott and members of the Centre for Gravitational Physics, where this work was completed, for their kind hospitality. This work was supported by the Academic Research Fund (WBS No.: R-144-000-277-112) from the National University of Singapore.

\bigskip\bigskip

{\renewcommand{\Large}{\normalsize}
}

\end{document}